\documentclass[onecolumn,floats,prb,aps,showpacs]{revtex4}
\usepackage{graphicx}
\usepackage{amsfonts}
\usepackage{amsmath}
\usepackage{amssymb}
\usepackage{exscale}
\usepackage{color}

\begin{document}

\title{Geometry-induced localization of thermal fluctuations in ultrathin
superconducting structures}
\author{W. V. Pogosov$^{1,2}$, V. R. Misko$^{1}$, and F. M. Peeters$^{1}$}
\affiliation{$^{1}$Departement Fysica, Universiteit Antwerpen, Groenenborgerlaan 171,
B-2020 Antwerpen, Belgium}
\affiliation{$^{2}$Institute for Theoretical and Applied Electrodynamics, Russian Academy
of Sciences, Izhorskaya 13, 125412, Moscow, Russia}
\date{\today }

\begin{abstract}
Thermal fluctuations of the order parameter in an ultrathin triangular
shaped superconducting structure are studied near $T_{c}$, in zero applied
field. We find that the order parameter is prone to much larger fluctuations
in the corners of the structure as compared to its interior. This
geometry-induced localization of thermal fluctuations is attributed to the
fact that condensate confinement in the corners is characterised by a lower
effective dimensionality, which favors stronger fluctuations.
\end{abstract}

\maketitle

\section{Introduction}

Recent progress in growth technology has made it possible to fabricate
ultrathin films, consisting of just few monoatomic layers. Very recently\cite%
{Pekin} superconducting properties of single-atomic layer films made of Pb
were reported, including even the observation of Abrikosov vortices. It is
also possible to make nano- and microstructures based on such thin films.
These nanostructures are generally of much higher quality than those, used
in earlier experiments\cite{exper}, which, in particular, were not
completely free of pinning centers.

It is well known that the lower the dimensionality of the system the
stronger the thermal fluctuations. Ultrathin superconducting films as well
as nanostructures based on such films are obvious candidates for the
observation of thermal fluctuations. Indeed, it was reported very recently
that strong current-induced thermal fluctuations of the order parameter were
observed in superconducting nanowires fabricated in a meander of NbN\cite%
{Bartolf}. Structures of this kind are used in highly-sensitive
photodetectors. It turns out that the fluctuations are a main source of
dark-count events in such photodetectors: this means that their effect is
actually parasitic. Fluctuation phenomena were interpreted in terms of a
thermally-activated entry of vortices, as well as by unbinding of
vortex-antivortex pairs\cite{Bartolf}. In another recent experiment\cite%
{Cren}, no hysteresis for vortex penetration and expulsion was observed in
the case of a nanoisland of Pb, which was so small that it could accommodate
only one vortex. In a similar experiment\cite{Nishio}, performed at lower
temperature, some hysteresis was detected, but the width of the hysteresis
region was significantly smaller than expected from theory. A theoretical
explanation for these phenomena was very recently suggested by one of us in
terms of a thermal suppression of the surface barrier for vortex entry/exit,
which might occur in superconducting nanoislands made of Pb in the regime of
the ultimate single vortex confinement\cite{Pogosov}. Previously, thermal
activation of vortices over the surface barrier was demonstrated to be
possible in high-$T_{c}$ low-dimensional structures\cite{Hern}.

Geometry itself can play an important role in fluctuation phenomena. The
shape of islands, prepared by the method of evaporation in ultrahigh vacuum,
depends strongly on their sizes. They grow according to the
Stranski-Krastanov scenario, i.e., starting from well-facetted nuclei\cite%
{Cren}. As a result, islands with lateral dimensions of nearly 100 nm and
smaller have a hexagonal shape, while larger islands tend to have a
triangular shape [see, e.g., Fig. 1(a) in Ref. \cite{Cren}]. At the same
time, it is known that the triangular geometry can lead to very peculiar
consequences for the superconducting condensate confined in this geometry.
For instance, it was shown in Refs. \cite{VVM,Misko} that stable vortex-%
\textit{antivortex} molecules can nucleate, when a homogeneous magnetic
field is applied. The aim of the present paper is to explore how
triangularity, or more generally the presence of sharp corners in a thin
superconducting structure, influences \textit{thermal fluctuations} of the
order parameter. The motivation to study the triangular geometry, apart from
possible links with experiments, is that one can expect thermal fluctuations
to be stronger in the corners of a triangular structure as compared to its
interior, since the superconducting condensate in the corners is strongly
confined by samples's borders. This implies that the system somehow is
characterized by lower effective dimensionality than in the interior. It is
however not evident a priori if this rather general argument leads to any
noticeable effect for the fluctuations of the order parameter in a triangle,
so that more careful inspection is certainly desirable. In order to reveal
the effect of geometry on spatial localization of thermal fluctuations, we
restrict ourselves to the case of zero applied field, but we consider
temperatures both below and above $T_{c}$. We apply the method of
small-amplitude oscillations within the Ginzburg-Landau theory, and we do
not consider the possible excitation of vortices.

The paper is organized as follows. In Section II we formulate our model. In
Section III we present our results for several correlation functions both
below and above $T_{c}$. We conclude in Section IV.

\section{Model}

The derivation presented in this Section applies for the case of
temperatures below $T_{c}$. It is straightforward to adopt it for the case
of temperatures above $T_{c}$.

\subsection{General formulation}

We start with the dimensionless Ginzburg-Landau functional for the
superconducting energy of the island of thickness $d$, in zero applied field:

\begin{equation}
F=\frac{B_{c}(T)^{2}}{\mu _{0}}\xi (T)^{2}d\int d^{2}r\left( -\left\vert
f\right\vert ^{2}+\frac{1}{2}\left\vert f\right\vert ^{4}+\left\vert \mathbf{%
\nabla }f\right\vert ^{2}\right) ,  \label{1}
\end{equation}%
where integration is performed over the cross section of the nanostructure, $%
f$ is the dimensionless order parameter. All distances are measured in units
of the temperature-dependent coherence length $\xi (T)$, and $B_{c}(T)$ is
the thermodynamical critical field given by%
\begin{equation}
B_{c}(T)=\frac{\Phi _{0}}{2\pi \sqrt{2}\xi (T)\lambda (T)}.  \label{2}
\end{equation}%
We consider ultrathin islands, with $d\ll \xi (T)$, so that the problem is
two-dimensional. The boundary condition for the order parameter at each of
the three edges of a sample is taken in its usual form for the case of a
superconductor/vacuum interface%
\begin{equation}
\frac{\partial f}{\partial \mathbf{n}}=0,  \label{3}
\end{equation}%
where $\mathbf{n}$\ is the unit vector in the direction perpendicular to the
edge.

Let us now estimate the ratio $G(T)$ of the energy needed to suppress the
order parameter to zero within the volume $\sim \xi (T)^{2}d$ and the
thermal energy $k_{B}T$, using realistic values for all parameters for
typical ultrathin nanostructures made of Pb that have been realized in
recent experiments. We define $G(T)$\ as%
\begin{equation}
G(T)=\frac{B_{c}(T)^{2}\xi (T)^{2}d}{2\mu _{0}k_{B}T}.  \label{4}
\end{equation}%
For the penetration depth, we use the usual expression\cite{Tinkham} for
dirty superconductors, $\lambda (T)\simeq 0.62\lambda _{0}\sqrt{\frac{\xi
_{0}/l}{1-T/T_{c}}}$ where $\lambda _{0}\simeq 40$~nm is the penetration
depth in bulk Pb, $l\approx 2d$ is the quasiparticle mean free part, and $%
T_{c}=7.2$ K. Then, for a film of thickness 2 nm (similar to Ref. \cite%
{Nishio} and nearly 3 times larger than in Ref. \cite{Cren}), operated at $%
T=0.75T_{c}$, we estimate $G(T)\sim 10$.

\subsection{Fluctuative modes}

Next, we expand the order parameter $f$ around its mean value in the absence
of fluctuations%
\begin{equation}
f=1+\delta f.  \label{5}
\end{equation}%
After substituting Eq. (5) into Eq. (1), we obtain the expression for the
increase of superconducting energy due to thermal fluctuations%
\begin{equation}
\delta F=\frac{B_{c}(T)^{2}}{\mu _{0}}\xi (T)^{2}d\int d^{2}r\left(
\left\vert \delta f\right\vert ^{2}+\frac{1}{2}(\delta f^{2}+\delta f^{\ast
2})+\left\vert \mathbf{\nabla }\delta f\right\vert ^{2}\right) ,  \label{6}
\end{equation}%
where we have kept only terms up to quadratic ones in $\delta f$ and $\delta
f^{\ast }$.

\begin{figure}[tbp]
\begin{center}
\includegraphics[width=0.5\textwidth]{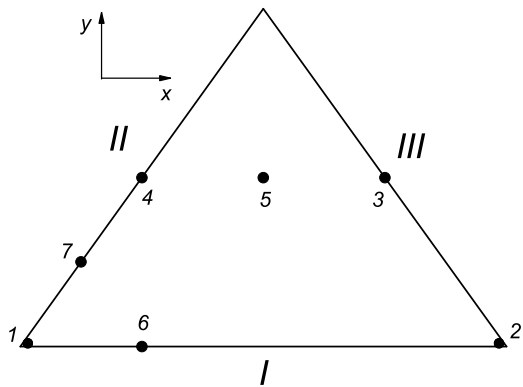}
\end{center}
\caption{ Schematic image of a triangular structure. Numbers depict points,
the correlation functions being calculated along the straight lines
connecting these points [see in the text].}
\label{Fig1}
\end{figure}

Obviously, $\delta f$ can be represented as a superposition of plane waves.
However, $\delta f$ for each fluctuative mode must satisfy the boundary
condition given by Eq. (3), which is a very strong restriction imposed by
the geometry of our problem. In order to circumvent this difficulty, we will
use, with certain modifications, an approach applied very recently in Ref. 
\cite{Sasha} for the triangular dot of graphene, for which exact electronic
wave functions were found. The major difference with the graphene dot
problem is in the boundary condition, which is vanishing of the wave
function at the triangle's border, instead of the vanishing of its first
derivative [Eq. (3)]. Following Ref. \cite{Sasha}, as a first step we focus
on the sector, confined between edges I and II [see Fig. 1] and we find the
form of $\delta f$, which satisfies boundary condition [Eq. (3)] at these
two edges. Consider a plane wave with the wave vector $\mathbf{k}%
_{1}=(k_{x},k_{y})$: 
\begin{equation}
\psi _{1}=\exp (-i\mathbf{k}_{1}\mathbf{r}).  \label{7}
\end{equation}%
When reflected from edge I, it is converted into wave $\psi _{2}$ with $%
\mathbf{k}_{2}=(k_{x},-k_{y})$. It is then easy to see that the sum of $\psi
_{1}$ and $\psi _{2}$ satisfies Eq. (3) at the edge I. This is in contrast
with Ref. \cite{Sasha}, where the difference between $\psi _{1}$ and $\psi
_{2}$ was taken due to the different boundary condition. When these two
waves are reflected from edge II, two new waves appear with $\mathbf{k}_{5}=-%
\frac{1}{2}(k_{x}+\sqrt{3}k_{y},-\sqrt{3}k_{x}+k_{y})$ and $\mathbf{k}_{6}=-%
\frac{1}{2}(k_{x}-\sqrt{3}k_{y},-\sqrt{3}k_{x}-k_{y})$. After the reflection
from edge I, these two ones give rise to two more waves with $\mathbf{k}%
_{3}=-\frac{1}{2}(k_{x}-\sqrt{3}k_{y},\sqrt{3}k_{x}+k_{y})$ and $\mathbf{k}%
_{4}=-\frac{1}{2}(k_{x}+\sqrt{3}k_{y},\sqrt{3}k_{x}-k_{y})$. Finally, after
reflecting from edge I, these last waves do not lead to any new waves. Then,
the \textit{sum} of the six wave functions $\psi _{1}+...+$ $\psi _{6}$
satisfies the boundary condition, given by Eq. (3), both at edges I and II,
whatever $\mathbf{k}_{1}$ is. The same boundary condition at edge III leads
to quantization rules for $(k_{x},k_{y})$, which differ from those found in
Ref. \cite{Sasha}. It is easy to see that the allowed $(k_{x},k_{y})$\ split
into two branches:%
\begin{eqnarray}
k_{x}^{(1)} &=&\frac{4\pi }{3L}m,  \label{8} \\
k_{y}^{(1)} &=&\frac{4\pi }{3L}n\sqrt{3},  \label{9}
\end{eqnarray}%
and%
\begin{eqnarray}
k_{x}^{(2)} &=&\frac{4\pi }{3L}\left( m+\frac{1}{2}\right) ,  \label{10} \\
k_{y}^{(2)} &=&\frac{4\pi }{3L}\left( n+\frac{1}{2}\right) \sqrt{3},
\label{11}
\end{eqnarray}%
where $L$ is the dimensionless length of the triangle's edge, while $m$ and $%
n$ are arbitrary integer numbers.

For these two branches, eigen energies are given by%
\begin{eqnarray}
E_{mn}^{(1)} &=&\left( \frac{4\pi }{3L}\right) ^{2}\left(
m^{2}+3n^{2}\right) ,  \label{12} \\
E_{mn}^{(2)} &=&\left( \frac{4\pi }{3L}\right) ^{2}\left[ \left( m+\frac{1}{2%
}\right) ^{2}+3\left( n+\frac{1}{2}\right) ^{2}\right] .  \label{13}
\end{eqnarray}%
Wave functions corresponding to different ($m$, $n$) but the same branch are
not necessarily unique. By a proper choice of $m$ and $n$, we obtain the
following basis of normalized and unique wave functions%
\begin{eqnarray}
\varphi _{mn}^{(1)} &=&\frac{1}{\sqrt{6}}\left( \psi _{1}^{(1)}+...+\psi
_{6}^{(1)}\right) ,\text{ }n>\left\vert m\right\vert ,  \label{14} \\
\varphi _{mn}^{(1)} &=&1,\text{ }n=m=0,  \label{15}
\end{eqnarray}%
where $\psi _{1}^{(1)}$,..., $\psi _{6}^{(1)}$ depend on ($k_{x}^{(1)}$,$%
k_{y}^{(1)}$), and 
\begin{equation}
\varphi _{mn}^{(2)}=\frac{1}{\sqrt{6}}\left( \psi _{1}^{(2)}+...+\psi
_{6}^{(2)}\right) ,\text{ }n>\left\vert m+1/2\right\vert ,  \label{16}
\end{equation}%
where $\psi _{1}^{(2)}$,..., $\psi _{6}^{(2)}$ depend on ($k_{x}^{(2)}$,$%
k_{y}^{(2)}$).

We can expand the fluctuating contribution to the order parameter in the
constructed basis:%
\begin{equation}
\delta f=\sum_{n,m,\alpha }c_{mn}^{(\alpha )}\varphi _{mn}^{(\alpha )},
\label{17}
\end{equation}%
where $\alpha =1,2$\ stand for the two branches. We now substitute expansion
(17) in Eq. (6). After performing the integration over the triangle's area,
we arrive at the following expression for the statistical sum%
\begin{equation*}
Z=\prod\limits_{n,m,\alpha }\int d(Re(c_{mn}^{(\alpha
)}))d(Im(c_{mn}^{(\alpha )}))\exp \left\{ -G(T)\frac{L^{2}\sqrt{3}}{4}\left[
(Re(c_{mn}^{(\alpha )}))^{2}(E_{mn}^{(\alpha )}+2)+(Im(c_{mn}^{(\alpha
)}))^{2}E_{mn}^{(\alpha )}\right] \right\} .
\end{equation*}%
It contains a product of Gaussian integrals, which can be easily evaluated
analytically as%
\begin{equation}
Z=\prod\limits_{n,m,\alpha }\frac{2\pi }{L\sqrt{E_{mn}^{(\alpha
)}(E_{mn}^{(\alpha )}+2)G(T)\sqrt{3}}}.  \label{18}
\end{equation}

\subsection{Correlation functions}

In order to study the spatial localization of fluctuations in the triangle
below $T_{c}$, we analyze separately phase and density fluctuations, since
they behave in different ways. Namely, these are fluctuations of the phase
of the order parameter which are responsible for the loss of the long-range
order\cite{Varlamov}.

It is straightforward to see, from Eq. (5), that the phase $\chi $ of the
order parameter can be expressed as%
\begin{equation}
\chi =\frac{1}{2i}\left( \delta f-\delta f^{\ast }\right) .  \label{19}
\end{equation}%
The quantity we are interested in is the correlation function%
\begin{equation}
K_{ph}(\mathbf{r}_{1},\mathbf{r}_{2})=\left\langle \left[ \chi (\mathbf{r}%
_{1})-\chi (\mathbf{r}_{2})\right] ^{2}\right\rangle _{T}.  \label{20}
\end{equation}%
In an infinite isotropic system this quantity depends only on the distance $%
\left\vert \mathbf{r}_{2}-\mathbf{r}_{1}\right\vert $. It diverges
logarithmically at large distances for two-dimensional systems, expressing
the loss of long-range order. However for a confined system with anisotropic
geometry, $K_{ph}(\mathbf{r}_{1},\mathbf{r}_{2})$ depends on both $\mathbf{r}%
_{1}$ and $\mathbf{r}_{2}-\mathbf{r}_{1}$.

Using Eqs. (17) and (19), we arrive at the following expression for $K_{ph}(%
\mathbf{r}_{1},\mathbf{r}_{2})$ 
\begin{eqnarray}
K_{ph}(\mathbf{r}_{1},\mathbf{r}_{2}) &=&\frac{1}{2}\sum_{m,n,\alpha
}\left\langle (Im(c_{mn}^{(\alpha )}))^{2}\right\rangle _{T}\left\{ Re\left[
\left( \Gamma _{mn}^{(\alpha )}(\mathbf{r}_{1},\mathbf{r}_{2})\right) ^{2}%
\right] +\left\vert \Gamma _{mn}^{(\alpha )}(\mathbf{r}_{1},\mathbf{r}%
_{2})\right\vert ^{2}\right\}  \notag \\
&&-\frac{1}{2}\sum_{m,n,\alpha }\left\langle (Re(c_{mn}^{(\alpha
)}))^{2}\right\rangle _{T}\left\{ Re\left[ \left( \Gamma _{mn}^{(\alpha )}(%
\mathbf{r}_{1},\mathbf{r}_{2})\right) ^{2}\right] -\left\vert \Gamma
_{mn}^{(\alpha )}(\mathbf{r}_{1},\mathbf{r}_{2})\right\vert ^{2}\right\} ,
\label{21}
\end{eqnarray}%
where%
\begin{equation}
\Gamma _{mn}^{(\alpha )}(\mathbf{r}_{1},\mathbf{r}_{2})=\varphi
_{mn}^{(\alpha )}(\mathbf{r}_{1})-\varphi _{mn}^{(\alpha )}(\mathbf{r}_{2}).
\label{22}
\end{equation}%
Thermally-averaged $(Re(c_{mn}^{(\alpha )}))^{2}$ and $(Im(c_{mn}^{(\alpha
)}))^{2}$\ can be found analytically as%
\begin{equation}
\left\langle (Re(c_{mn}^{(\alpha )}))^{2}\right\rangle _{T}=\frac{2}{%
G(T)L^{2}\sqrt{3}\left( E_{mn}^{(\alpha )}+2\right) },  \label{23}
\end{equation}%
\begin{equation}
\left\langle (Im(c_{mn}^{(\alpha )}))^{2}\right\rangle _{T}=\frac{2}{%
G(T)L^{2}\sqrt{3}E_{mn}^{(\alpha )}}.  \label{24}
\end{equation}%
By substituting Eqs. (23) and (24) into Eq. (21) and performing the
summations, we calculate $K_{ph}(\mathbf{r}_{1},\mathbf{r}_{2})$\ along
various lines inside the triangle. This sum, however, is divergent at large $%
m$ and $n$, and this divergency has to be cut in a standard way [see, e.g.,
p. 336 of Ref. \cite{Varlamov}] at wave vectors $k\sim 1/\xi (T)$. We found
that the dependence of the final results on the particular choice of the
cutoff value is rather weak, within 10 \%.

In a similar way, we can express the fluctuation part $\delta n_{s}$ of the
density of superconducting electrons $\left\vert f\right\vert ^{2}$ in terms
of $\delta f$ and $\delta f^{\ast }$%
\begin{equation}
\delta n_{s}=\delta f+\delta f^{\ast }.  \label{25}
\end{equation}%
By using the developed approach, it is also possible to find the
density-density correlation function, defined as 
\begin{equation}
K_{den}(\mathbf{r}_{1},\mathbf{r}_{2})=\left\langle \delta n_{s}(\mathbf{r}%
_{1})\delta n_{s}(\mathbf{r}_{2})\right\rangle _{T}.  \label{26}
\end{equation}

Thermal fluctuations of the order parameter also occur at temperatures above 
$T_{c}$. In this case, the equilibrium order parameter is equal to zero,
which leads to changes in Eqs. (5), (18) and therefore also in Eqs. (23) and
(24). We omit the derivation and we will present our final results in
Section III for the correlation function for the order parameter itself,
without separating the density and the phase 
\begin{equation}
K(\mathbf{r}_{1},\mathbf{r}_{2})=\left\langle f(\mathbf{r}_{1})^{\ast }f(%
\mathbf{r}_{2})\right\rangle _{T}.  \label{27}
\end{equation}%
Above $T_{c}$, it is also convenient to normalize all the distances by $\xi
(T)$, where $\xi (T)=\sqrt{\hbar ^{2}/2m\left\vert \alpha (T)\right\vert }$.

\section{Results and discussion}

In this Section, we discuss the behavior of the correlation functions in the
triangle both below and above $T_{c}$.

\subsection{Temperatures below $T_{c}$}

\begin{figure}[tbp]
\begin{center}
\includegraphics[width=0.5\textwidth]{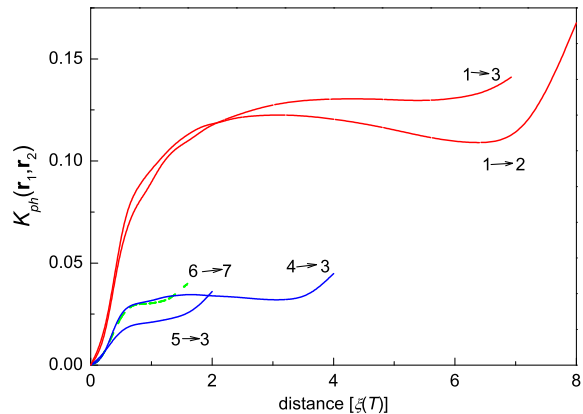}
\end{center}
\caption{ (Color online) Dependence of the correlation function $K_{ph}(%
\mathbf{r}_{1},\mathbf{r}_{2})$ on $\left\vert \mathbf{r}_{2}-\mathbf{r}%
_{1}\right\vert $ along different paths in the triangular structure below $%
T_{c}$. Red lines show $K_{ph}(\mathbf{r}_{1},\mathbf{r}_{2})$ along the
paths, which originate from the corner. Blue lines correspond to paths in
the interior of the structure. Green dashed line corresponds to the path
across the bisector.}
\label{Fig2}
\end{figure}

Our results for the dependence of the phase correlation function $K_{ph}(%
\mathbf{r}_{1},\mathbf{r}_{2})$ on $\left\vert \mathbf{r}_{2}-\mathbf{r}%
_{1}\right\vert $ along different paths are shown in Fig. 2 for a triangle
with $L=8$ [several hundreds of nanometers for the case of a nanostructure
made of Pb at $T=0.75T_{c}$, $G(T=0.75T_{c})=10$]. Different curves show the
behavior of $K_{ph}(\mathbf{r}_{1},\mathbf{r}_{2})$ along straight lines
connecting various points inside the triangle, which are depicted in Fig. 1.
Curve $1\rightarrow 2$ corresponds to the path between the two corners of
the structure. Curve $1\rightarrow 3$ gives $K_{ph}(\mathbf{r}_{1},\mathbf{r}%
_{2})$ along the bisector starting from the corner. Curves $4\rightarrow 3$
and $5\rightarrow 3$ correspond to lines in the interior region of the
structure [lengths of segments $1-4$ and $2-3$ are $L/2$, while lengths of
segments $4-5$ and $5-3$ are the same]. Curve $6\rightarrow 7$ shows $K_{ph}(%
\mathbf{r}_{1},\mathbf{r}_{2})$ across the bisector [lengths of segments $%
1-6 $ and $1-7$ are $L/5$].

\begin{figure}[tbp]
\begin{center}
\includegraphics[width=0.5\textwidth]{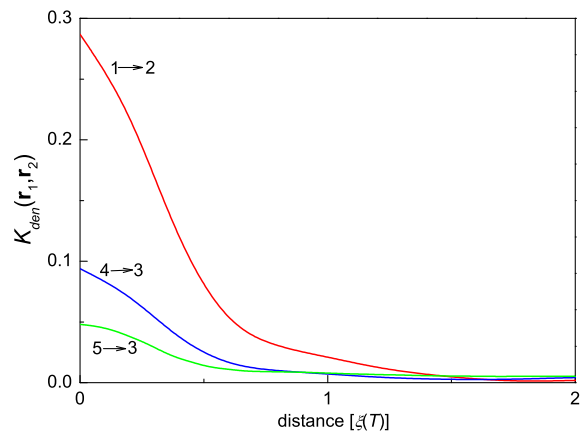}
\end{center}
\caption{ (Color online) Dependence of the correlation function $K_{den}(%
\mathbf{r}_{1},\mathbf{r}_{2})$ on $\left\vert \mathbf{r}_{2}-\mathbf{r}%
_{1}\right\vert $ along different paths within the triangular structure
below $T_{c}$. Red line shows $K_{den}(\mathbf{r}_{1},\mathbf{r}_{2})$ along
the path, which originates from the corner. Blue line corresponds to path,
which starts from the edge towards the interior of the structure. Green line
shows $K_{den}(\mathbf{r}_{1},\mathbf{r}_{2})$ starting from the interior to
the edge.}
\label{Fig3}
\end{figure}

The shape of all these curves is similar and it generally resembles the
behavior of the same correlation function for an isotropic infinite
two-dimensional system. Namely, it first grows rapidly with increasing $%
\left\vert \mathbf{r}_{2}-\mathbf{r}_{1}\right\vert $ and then starts to
demonstrate a smoother behavior. There are, however, obvious differences
between the curves, both quantitative and qualitative. First, it is seen
that the initial increase of $K_{ph}(\mathbf{r}_{1},\mathbf{r}_{2})$ is much
larger [several times] for those paths which originate from the corner
[curves $1\rightarrow 2$ and $1\rightarrow 3$], where, according to our
initial guess, fluctuations are stronger. Thus, we can conclude that this
guess is justified. Second, $K_{ph}(\mathbf{r}_{1},\mathbf{r}_{2})$ along
the curve $1\rightarrow 2$ starts to rapidly increase again, when
approaching the other corner, while the such an increase for other curves is
not so pronounced. This feature implies that the order parameter between 
\textit{different} corners is more decoherent, which again supports our
hypothesis. For the range of parameters used here, we see that the maximum $%
\sqrt{K_{ph}(\mathbf{r}_{1},\mathbf{r}_{2})}$ is nearly 0.4, which implies
that the coherence between the order parameter inside different corners
starts to be lost.

Figure 3 gives the density-density correlation function $K_{den}(\mathbf{r}%
_{1},\mathbf{r}_{2})$ for three directions, along paths $1\rightarrow 2$, $%
4\rightarrow 3$, and $5\rightarrow 3$. We again see that the correlation
function in the corner [$1\rightarrow 2$] is much larger than the same
function in the interior [$4\rightarrow 3$ and $5\rightarrow 3$]. This shows
that not only the phase of the order parameter, but also the density of
superconducting electrons fluctuates stronger in the corners. It is also of
interest to note that the comparison of $K_{den}(\mathbf{r}_{1},\mathbf{r}%
_{2})$ along paths $4\rightarrow 3$ and $5\rightarrow 3$\ shows that
fluctuations are stronger near the triangle's surface as compared to the
central part of the nanostructure.

The strength of thermal fluctuations at the given point $\mathbf{r}_{1}$ may
be characterized by the density-density correlation function $K_{den}(%
\mathbf{r}_{1},\mathbf{r}_{2})$ with $\left\vert \mathbf{r}_{2}-\mathbf{r}%
_{1}\right\vert \sim \xi (T)$ averaged over all such values of $\mathbf{r}%
_{2}$\cite{Varlamov,Shmidt}. Although $K_{den}(\mathbf{r}_{1},\mathbf{r}_{2})
$ is usually peaked within nearly $1.5\xi (T)$ in the corners for various
triangle's sizes, such an averaging implies that the typical width of the
geometry-induced fluctuating region in each of the three corners is broader:
we estimate it to be several $\xi (T)$.

\begin{figure}[tbp]
\begin{center}
\includegraphics[width=0.5\textwidth]{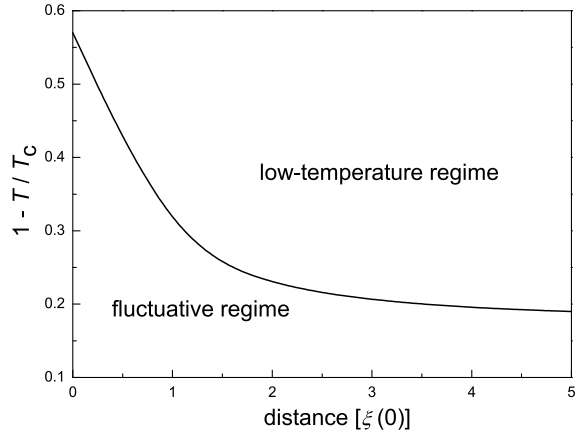}
\end{center}
\caption{ "Phase diagram" of the triangular structure below $T_{c}$ indicating
the region where fluctuations are more pronounced.}
\label{Fig4}
\end{figure}

In order to visualize the effect of geometry on thermal fluctuations, we
construct a "phase diagram" for the case of a triangular Pb nanostructure
with the edge length $15\xi (0)$ and thickness $d=5$ nm. We take $%
1\rightarrow 3$ path [along the bisector] and for each point $\mathbf{r}$ at
this path we calculate $\sqrt{K_{den}(\mathbf{r},\mathbf{r})}$, which shows
the average fluctuation of the density of superconducting electrons. We have
chosen the following qualitative criterion: if this quantity exceeds 0.15 at
a certain temperature, we enter the fluctuative regime. Our results for the
crossover between low-temperature and fluctuative regimes are presented in
Fig. 4 in the plane of the distance along the $1\rightarrow 3$ path and $%
1-T/T_{c}$. We indeed see that the characteristic temperature for the
crossover in the corners strongly depends on the position. Of course, the
transition between the two domains in Fig. 4 is not abrupt. It is more
appropriate to talk of different regimes, which are separated by a crossover
region. Thus, the suggested picture is very qualitative and it is given for
illustrative purposes to supplement an idea we want to convey.

\subsection{Temperatures above $T_{c}$}

\begin{figure}[tbp]
\begin{center}
\includegraphics[width=0.5\textwidth]{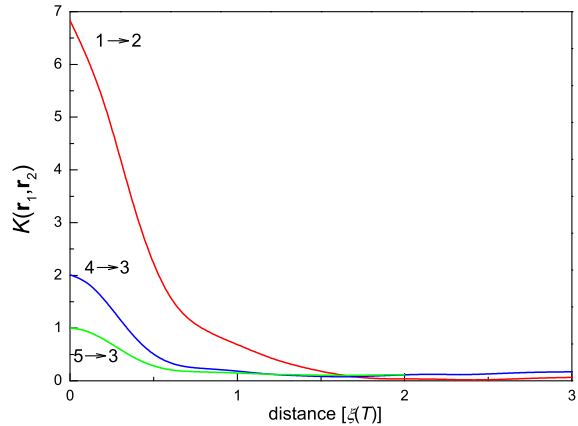}
\end{center}
\caption{ (Color online) Dependence of the correlation function $K(\mathbf{r}%
_{1},\mathbf{r}_{2})$ on $\left\vert \mathbf{r}_{2}-\mathbf{r}%
_{1}\right\vert $ along different paths within the triangular structure
above $T_{c}$. Red line shows $K(\mathbf{r}_{1},\mathbf{r}_{2})$ along the
path, which originates from the corner. Blue line corresponds to path, which
starts from the edge towards the interior of the structure. Green line shows 
$K(\mathbf{r}_{1},\mathbf{r}_{2})$ starting from the interior to the edge.}
\label{Fig5}
\end{figure}

The order parameter correlation function $K(\mathbf{r}_{1},\mathbf{r}_{2})$
above $T_{c}$, defined in Eq. (27), has to decrease\cite{Varlamov} with
increasing $\left\vert \mathbf{r}_{2}-\mathbf{r}_{1}\right\vert $. The
results of our calculations for $K(\mathbf{r}_{1},\mathbf{r}_{2})$ along
different paths are presented in Fig. 5, where the correlation function is
plotted for $T=1.25T_{c}$ for the triangle with length $L=8$\ [in terms of $%
\xi (T)$]. $K(\mathbf{r}_{1},\mathbf{r}_{2})$ is normalized by its value $K(%
\mathbf{r},\mathbf{r})$ in the point 5, which is located in the central part
of the nanostructure ["bulk"]. We indeed see the expected rapid decay of $K(%
\mathbf{r}_{1},\mathbf{r}_{2})$ with increasing $\left\vert \mathbf{r}_{2}-%
\mathbf{r}_{1}\right\vert $, but again the correlation function calculated
along the path, which originates from the corner [$1\rightarrow 2$], is
rather different from those, which correspond to lines inside the interior
of the structure. Namely, the average value of the density of
superconducting electrons induced by thermal fluctuations is much larger in
the corners. Fluctuations near the surface are stronger than far from the
surface [curves $4\rightarrow 3$ and $5\rightarrow 3$]. The width of the
geometry-enhanced fluctuating region in the corners is again estimated to be
several $\xi (T)$.

The comparison of our results for the correlation functions below and above $%
T_{c}$ shows that superconductivity is getting suppressed, below $T_{c}$,
starting from the corners, while, above $T_{c}$, the order parameter
preferentially nucleates again in the corners. However, there is no
contradiction, since if we consider a triangle with \textit{fixed} sizes and
if we start to increase $T$ towards $T_{c}$, $\xi (T)$\ becomes infinitely
large at $T_{c}$. This means that, close enough to $T_{c}$ on both sides of
the transition, the triangle is actually in the zero-dimensional regime, so
that the order parameter does not vary in space. In other words, the
fluctuating region in the vicinity of $T_{c}$ expands from the corners to
the whole structure, while by tuning temperature away from $T_{c}$, one can
make thermal fluctuations stronger in the corners than in the interior. Of
course, the length of the triangle's edge should be much larger than $\xi
(0) $ in order to see such a spatial localization of thermal fluctuations.

The geometry-induced localization of thermal fluctuations is expected to
occur in nanostructures of various shapes, not only triangular ones. The
general tendency is obvious: the sharper the corner, the stronger the
fluctuations in this corner. For instance, in the corners of square-shaped
samples, the width of the fluctuative region can be expected to be much
smaller than in the corners of the triangular structures.

\section{Conclusion}

We have studied thermal fluctuations of the order parameter in
quasi-two-dimensional superconducting structures of triangular shape. We
considered the case of zero applied field both below and above $T_{c}$. It
was shown that the order parameter exhibits much larger fluctuations in the
corners of the structure and the width of such a geometry-induced
fluctuating region is several $\xi (T)$. This unusual behavior can be
attributed to the fact that the confinement of the condensate in the corners
lowers locally the effective dimensionality of the system, thus making
thermal fluctuations more pronounced. The condensate in the corners can
serve as a source of thermal noise in superconducting devices based on
ultrathin nano- and microstructures.

\section{Acknowledgements}

This work was supported by the \textquotedblleft Odysseus\textquotedblright\
Program of the Flemish government, FWO-Vl, and the Belgian Science Policy
(IAP). W.V.P. acknowledges supports from the RFBR [Project no. 09-02-00248]
and the "Dynasty Foundation".


\begin{thebibliography}{99}
\bibitem{Pekin} T. Zhang, P. Cheng, W.-J. Li, Y.-J. Sun, G. Wang, X.-G. Zhu,
K. He, L. Wang, X. Ma, X. Chen, Y. Wang, Y. Liu, H.-Q. Lin, J.-F. Jia, and
Q.-K. Xue, Nature Phys. \textbf{6}, 104 (2010).

\bibitem{exper} A. K. Geim, S. V. Dubonos, I. V. Grigorieva, K. S.
Novoselov, F. M. Peeters, and V. A. Schweigert, Nature (London) \textbf{407}%
, 55 (2000).

\bibitem{Bartolf} H. Bartolf, A. Engel, A. Schilling, K. Il'in, M. Siegel,
H.-W. Hubers, and A. Semenov, Phys. Rev. B \textbf{81}, 024502 (2010).

\bibitem{Cren} T. Cren, D. Fokin, F. Debontridder, V. Dubost, and D.
Roditchev, Phys. Rev. Lett. \textbf{102}, 127005 (2009).

\bibitem{Nishio} T. Nishio, T. An, A. Nomura, K. Miyachi, T. Eguchi, H.
Sakata, S. Lin, N. Hayashi, N. Nakai, M. Machida, and Y. Hasegawa, Phys.
Rev. Lett. \textbf{101}, 167001 (2008).

\bibitem{Pogosov} W. V. Pogosov, Phys. Rev. B \textbf{81}, 184517 (2010).

\bibitem{Hern} A. D. Hernandez, B. J. Baelus, D. Dominguez, and F. M.
Peeters, Phys. Rev. B \textbf{71}, 214524 (2005); J. R. Kirtley, C. C.
Tsuei, V. G. Kogan, J. R. Clem, H. Raffy, and Z. Z. Li, Phys. Rev. B \textbf{%
68}, 214505 (2003).

\bibitem{VVM} L. F. Chibotaru, A. Ceulemans, V. Bryndoncx, and V. V.
Moshchalkov, Nature (London) \textbf{408}, 833 (2000).

\bibitem{Misko} V. R. Misko, V. M. Fomin, J. T. Devreese, and V. V.
Moshchalkov, Phys. Rev. Lett. \textbf{90}, 147003 (2003).

\bibitem{Tinkham} M. Tinkham, \textit{Introduction to Superconductivity},
Dover Publications, New York (2004).

\bibitem{Sasha} A. V. Rozhkov and F. Nori, Phys. Rev. B \textbf{81}, 155401
(2010).

\bibitem{Varlamov} A. Larkin and A. Varlamov, \textit{Theory of Fluctuations
in Superconductors,} Oxford University Press, Oxford (2004).

\bibitem{Shmidt} V. V. Schmidt, \textit{The Physics of Superconductors.
Introduction to Fundamentals and Applications}, edited by P. Muller and A.
V. Ustinov, Springer-Verlag, Berlin (1997).
\end{thebibliography}
\end{document}